\documentclass[a4paper,11pt]{article}
\pdfoutput=1 

\usepackage{jcappub} 

\usepackage[T1]{fontenc} 

\title{\boldmath Constraints on GeV Dark Matter interaction with baryons, from a novel dewar experiment}

\author{Xingchen Xu}
\author{and Glennys R. Farrar}
\affiliation{Center for Cosmology and Particle Physics, New York University\\New York, NY, 10003, USA}

\emailAdd{xingchen.xu@nyu.edu}
\emailAdd{glennys.farrar@nyu.edu}

\abstract{Dark matter which scatters off ordinary baryonic matter with a relatively large cross section cannot be constrained by traditional, deep underground WIMP experiments, due to the energy loss of DM in the Earth's atmosphere and crust. However, for a sufficiently large cross section, DM particles in the GeV mass range can be captured and thermalized within Earth, resulting in the accumulation of a DM atmosphere whose number density can be as large as $10^{14} \text{ cm}^{-3}$ at Earth's surface.  (If the DM-nucleon interaction is attractive and bound states can be formed, most DM bind to nuclei and the density of the atmosphere would be much lower.)

Neufeld and Brach-Neufeld performed experiments to constrain the DM-baryon scattering cross section of DM particles forming an atmosphere around Earth, by measuring the evaporation rate of liquid nitrogen in a storage dewar within which various materials are immersed. If the DM-nitrogen cross section is in an appropriate range, room temperature DM would penetrate the dewar walls and scatter on the cold nitrogen, increasing its evaporation rate beyond the observed level. Limits on the cross section of DM with other materials than nitrogen are obtained by adding known amounts of different materials; if the material is heated by interactions with DM, that heats and evaporates the liquid nitrogen. 

Because Born approximation is in general invalid in much of the relevant cross section regime, and in fact gives dramatically incorrect results, it is non-trivial to interpret such experimental results as a limit on the DM-nucleon cross section. In this paper we derive the constraints from the Neufeld experiments on the parameter space for DM-baryon scattering, with the interaction modeled as a Yukawa potential sourced by the finite sized nucleus.  Combining the dewar constraints with BBN,  we exclude for the first time a cross section above $10^{-26} \text{ cm}^{2}$ for DM mass 0.8-5.5 GeV, for any sign interaction.  One DM model that is constrained is sexaquark $(uuddss)$ DM with mass $m_X \sim 2$ GeV; it remains viable.}

\begin{document} 
\maketitle
\flushbottom
\newpage

\section{\label{sec:introduction}Introduction}

The possiblity of an interaction between dark matter (DM) and standard model particles has been a driving force in the study of dark matter and new physics. In recent years, great improvements have been made in direct detection searches for WIMP dark matter ~\cite{Aprile:2018, PandaX-II:2020oim} and the limit on the spin-independent (SI) WIMP-nucleon cross section has been constrained to be smaller than about $10^{-46}\rm{\,\,cm^2}$ for dark matter with $m_X\sim$ 10 to 100 GeV. On the other hand, these deep underground experiments are generally insensitive to a cross section above  $\sim 10^{-30}\, \rm{cm^2}$ 
because DM particles lose too much energy due to scattering with Earth's atmosphere and crust before reaching the detector~\cite{Starkman:1990nj,Mahdawi:2017cxz,Mahdawi:2018euy}. One way to get rid of the overburden is to perform the experiment above Earth's surface and atmosphere. The XQC  sounding rocket experiment~\cite{McCammon:2002} has been used to place a limit in this cross section regime~\cite{Wandelt:2000ad,Zaharijas:2004jv,Erickcek:2007jv,Mahdawi:2018euy}. However the minimum DM mass for which XQC is sensitive is limited by the energy threshold of the detector, which is quite uncertain because the thermalization efficiency of the detector for low energy nuclear recoils has never been measured.  A ball-park estimate is that XQC may be sensitive down to $m_X\sim \rm{few\,GeV}$~\cite{Mahdawi:2018euy}. 

\begin{figure}
\centering 
\includegraphics[width=0.7\textwidth]{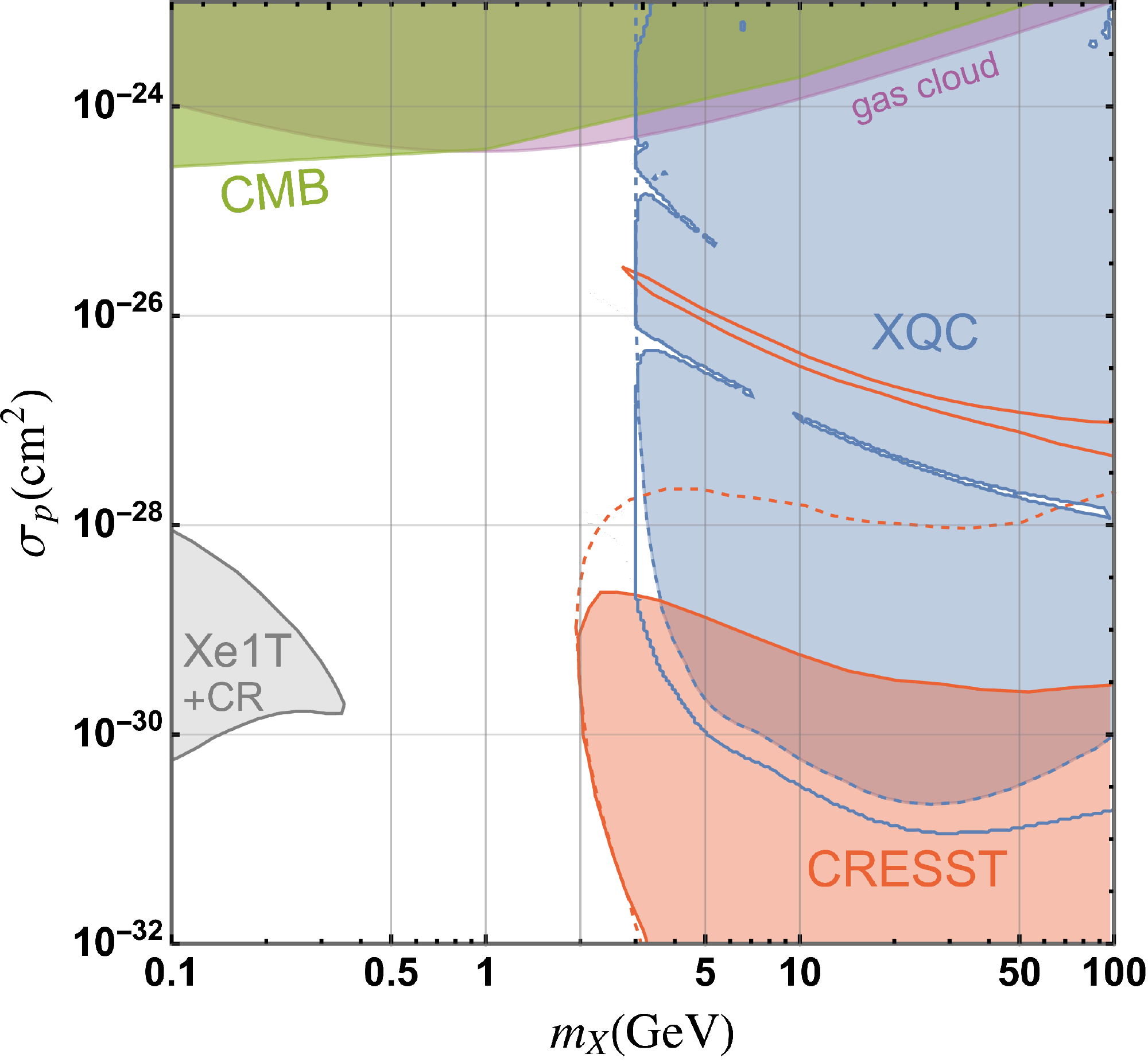}
\caption{\label{fig:experiments}Limits on DM-nucleon spin-independent cross section. CR (gray) is obtained based on cosmic ray up-scattering of DM~\cite{Bringmann:2018cvk}. For CMB (green)~\cite{Xu:2018efh}, gas cloud cooling (purple) ~\cite{Wadekar:2019mpc}, XQC (blue) ~\cite{Wandelt:2000ad,Zaharijas:2004jv,Erickcek:2007jv,Mahdawi:2018euy} and CRESST surface detector (red)~\cite{Angloher:2017sxg,Mahdawi:2018euy}, results obtained from the correct non-perturbative treatment in~\cite{Xu:2020qjk_v2} are shown (for mediator mass $m_\phi \gtrsim$ GeV ). For XQC and CRESST, colored region are conservatively excluded while solid and dashed lines indicate additional excluded region if the sign of the interaction is determined to be attractive (solid) or repulsive (dashed). We display the limit corresponding to a thermalization efficiency of 0.01; simple estimates suggest a higher efficiency is implausible~\cite{Mahdawi:2018euy,Farrar:2020}. 
}
\end{figure}

Robust cosmological and astrophysical constraints have also been obtained from CMB observations~\cite{Xu:2018efh} and from limits on anomalous heating of gas-rich dwarf galaxies~\cite{Wadekar:2019mpc}, which cover both the sub-GeV and above-GeV mass range with relatively large cross section bounds of order $10^{-25}\,{\rm cm}^2$.\footnote{ If use of the Lyman-$\alpha$ forest to place constraints on small scale structure were valid, the CMB limits could be improved to $\sim 10^{-26}\,{\rm cm}^2$.  But structure can form on the Lyman-$\alpha$ scale from non-cosmological sources, so these limits are not robust~\cite{Hui:2016ltb,vandenBosch+Shattering19}.  Similarly, recent limits using Milky Way dwarf galaxies~\cite{DES:2020fxi} may not be robust.  See~\cite{Xu:2018efh} for a recent comprehensive study, for the first time correctly treating non-perturbative effects which are important for the relatively large cross sections being probed and vitiate the simple Born-approximation scaling of cross sections with nuclear size, $A$.} Limits excluding DM with masses up to 0.3 GeV having a cross section on nucleons in the $10^{-28}- 10^{-30}\, {\rm cm}^2$ range been obtained by ~\cite{Bringmann:2018cvk} using Xenon1T based on up-scattering of DM by high energy cosmic rays.\footnote{We note for completeness that the SENSEI experiment is sensitive to a range of DM masses below the GeV scale, but only for DM which scatters on electrons.} Figure~\ref{fig:experiments} summarizes the robust constraints for DM mass in the 0.1-100 GeV and cross sections above a micro-barn.   The inadequacy of the constraints for GeV-scale dark matter is striking.  It is our aim in this paper to take steps toward probing this DM mass and cross section range.

Recently, it was shown by Neufeld, Farrar and McKee~\cite{Neufeld:2018slx} (NFM18 below) that 
dark matter particles can be captured by the Earth, thermalize, and produce a dark matter atmosphere for the Earth. Dark matter particles concentrated in this way can have a number density much higher than the Galactic average, reaching $10^{14} \rm{\,\, cm^{-3}}$ at Earth's surface for $m_X=(1-2)$ $m_p$ if $\sigma\gtrsim 10^{-28} \rm{\,\,cm^2}$ for a relevant combination of nuclei.  NFM18 placed constraints on DM-nucleus scattering cross section for this low energy, high concentration component of DM from four considerations: the LHC beam life time, the orbital decay of the Hubble Space Telescope (HST), the thermal conductivity of the Earth's crust and the vaporization rates of cryogenic liquids in insulated dewars. In a follow-up paper Neufeld and Brach-Neufeld~\cite{Neufeld:2019xes} (NBN19 below) performed a search for anomalous heating of samples of different materials containing different nuclei that are immersed in liquid nitrogen storage dewars. Any significant scattering of thermalized DM particles in their samples can result in a increased evaporation rate of the liquid nitrogen. 27 atomic nuclei were  included in the study and constraints on $n_A \sigma_A$ have been reported for all of them, with all together 74 different $A$ (including isotopes) ranging from 1 to 142 including the isotopes.  

Unfortunately, converting the NFM18 and NBN19 constraints on combinations of cross sections for different nuclei to limits on the  DM-nucleon cross section, as needed for constraining theoretical models, is highly non-trivial.  The cross sections probed are in a range in which $\sigma_A$ has a strong non-monotonic dependence on the atomic mass number $A$ of the target nucleus \cite{Xu:2019, Digman:2019,Farrar:2020,Xu:2020qjk_v2}.  It is highly non-perturbative and Born approximation is not applicable. 
However extracting the physics from the dewar limits is strongly motivated since the GeV mass range is otherwise very poorly constrained.  In particular, the sexaquark DM candidate~\cite{Farrar:2017eqq,Farrar:2020}, $uuddss$, falls in this mass and cross section regime.

In this paper we will show how to translate the limits on 74 different $n_A \sigma_A$ from NBN19 into constraints on parameter space for DM-baryon scattering modelled by a Yukawa potential, and finally obtain the combined constraint on $\sigma_p$, which covers the under-explored GeV dark matter region in Fig.~\ref{fig:experiments}. 

We begin by briefly summarizing in section~\ref{sec:NFMNB} the method and results of NFM18 for the Earth's DM halo and the limits on DM-baryon interaction from the dewar experiment of Neufeld and Brach-Neufeld in NBN19. 
In section~\ref{sec:Yukawa} we show how to interpret the results of the dewar experiment using a Yukawa potential as introduced in~\cite{Xu:2020qjk_v2}, and we put constraints on the parameter space ($\alpha, m_X$), where $\alpha$ is the DM-baryon coupling. 
In section~\ref{sec:cxlimits} we place constraints on the DM-nucleon cross section $\sigma_p$ and compare with limits obtained from other experiments and observations. Then we summarize our work in section~\ref{sec:summary}. Throughout the paper we use the mediator mass $m_\phi=1$ GeV as a benchmark; this is a natural value for the sexaquark application and at the end of section~\ref{sec:cxlimits} we discuss how the results change for different $m_\phi$.

\section{\label{sec:NFMNB}DM capture by Earth and Experimental Constraints}
Here we briefly summarize the DM capture scenario in NFM18 and the DM-baryon cross section limits in NFM18 and NBN19.  Readers are encouraged to refer the original papers for detailed information.

\subsection{\label{sec:NFM} DM capture by Earth}
Assuming Earth has an effective cross section $\pi R_{\oplus}^2$ with the halo DM, the total number of DM particles intercepted by the Earth during its lifetime can be approximated as
\begin{equation}
\label{eq:NI}
N_{\mathrm{I}}=\frac{t_{\oplus} \rho_{\mathrm{DM}} v_{\oplus} \pi R_{\oplus}^{2}}{m_{\mathrm{DM}}}
\end{equation}
where $t_{\oplus}=4.55 \rm{\,Gyr}$ is the age of the Earth, $\rho_{\mathrm{DM}}$ is the local DM mass density in the Galactic plane, $v_{\oplus}$ is the mean relative velocity between Earth and Galactic DM halo particles,  and $R_{\oplus}=6371\rm{\,km}$ is the radius of the Earth. 

Of course, not all DM particles intercepted by the Earth are captured and gravitationally bound since $v_{\oplus}$ is much greater than the escape velocity $v_{\rm{es}} = 11.2\rm{\,km/s}$. To become bound a DM particle must suffer enough collisions to reduce its velocity below $v_{\rm{es}}$ before it is reflected back into space by the crust and atmosphere. Since we are focusing on DM that has a hardronic cross section with baryons, the mean free path is much smaller than the Earth's radius so we do not need to consider DM that emerges at the other side of the Earth. NFM18 calculated the fraction of incident DM particles that are captured to be 
\begin{equation}\label{eq:fcap}
f_{\text {cap }}=2 \pi^{-1 / 2} N_{0}^{-1 / 2}=2 \pi^{-1 / 2}\left[\ln \left(v_{\mathrm{es}}^{2} / v_{\oplus}^{2}\right) / \ln \left(1-\bar{f}_{\mathrm{KE}}\right)\right]^{-1 / 2}
\end{equation}
where $N_{0}=\ln \left(v_{\mathrm{es}}^{2} / v_{\oplus}^{2}\right) / \ln \left(1-\bar{f}_{\mathrm{KE}}\right)$ is the smallest mean number of collisions for DM to be captured and 
\begin{equation}\label{fKE}
\bar{f}_{\mathrm{KE}}\equiv \frac{2 m_{\mathrm{DM}} m_{\mathrm{A}} }{\left(m_{\mathrm{DM}}+m_{\mathrm{A}}\right)^{2}}
\end{equation}
is the mean kinetic energy transfer for each collision.  For a fiducial DM mass value $m_X=2m_p$, and DM particles with velocity 200 km/s scattering in the atmosphere, $f_{\rm{cap}}=0.23$. Ignoring the loss of captured DM from the atmosphere and crust for now,  the average number density within the Earth's whole volume is
\begin{multline}\label{eq:nDM}
\bar{n}_{\mathrm{DM}}=\frac{3 f_{\mathrm{fap}} t_{\oplus} \rho_{\mathrm{DM}} v_{\oplus}}{4 R_{\oplus} m_{\mathrm{DM}}}\\=2.46 \times 10^{14}\left(\frac{f_{\mathrm{cap}}}{0.23}\right)\left(\frac{\rho_{\mathrm{DM}}}{0.3\left(\mathrm{GeV} / \mathrm{c}^{2}\right) \mathrm{cm}^{-3}}\right)\left(\frac{v_{\oplus}}{200 \mathrm{~km} \mathrm{~s}^{-1}}\right)\left(\frac{m_{\mathrm{DM}}}{m_{\mathrm{p}}}\right)^{-1} \mathrm{~cm}^{-3}.
\end{multline}
It is clear from Eq.~\eqref{eq:nDM} that the captured DM could greatly exceed the Galactic average, as much as $\sim 10^{14}$ larger for a  $m_X\sim$ GeV DM particle, which opens new channel for direct detection experiments if the DM number density at Earth's surface is also large. Assuming the DM particles are in thermal equilibrium with the local Earth's crust and atmosphere,  the DM density profile can be solved from the following equation:
\begin{equation}\label{eq:profileDIff}
\frac{d \ln p_{\mathrm{DM}}}{d r}=-\frac{m_{\mathrm{DM}}}{k T(r)} \frac{d \Phi}{d r}=-\frac{m_{\mathrm{DM}} g(r)}{k T(r)}=-\frac{G m_{\mathrm{DM}} M_{r}}{r^{2} k T(r)},
\end{equation}
where $p_{\mathrm{DM}}=n_{\mathrm{DM}} k T$ is the DM partial pressure, $T(r)$ is the local temperature and $M_r$ is the mass within radius r. From this the DM number density at Earth's surface, which is relevant for direct detection experiments, is found.  For example,  assuming $m_\mathrm{DM} = 2 m_\mathrm{p}$ for sexaquark DM~\cite{Farrar:2017eqq,Farrar:2020}, it can be obtained that $n_{\mathrm{DM}}\left(R_{\oplus}\right)=0.74 \bar{n}_{\mathrm{DM}}$.

Not all captured DM are retained. They can escape just like gas molecules in the upper atmosphere. Equation~\eqref{eq:nDM} for $\bar{n}_{\mathrm{DM}}$ must be modified to include the loss of DM. If the fractional loss rate is $f_{\rm{loss}}$,  then $\bar{n}_{\mathrm{DM}}$ will change with rate 
\begin{equation}\label{eq:dnbardt}
\frac{d \bar{n}_{\mathrm{DM}}}{d t}=\frac{3 f_{\text {cap }} \rho_{\mathrm{DM}} v_{\oplus} }{4 m_{\mathrm{DM}} R_{\oplus} }-\bar{n}_{\mathrm{DM}} f_{\mathrm{loss}},
\end{equation}
and achieve the final value today (assuming the capture and loss rate are roughly constant throughout the history of the Earth)
\begin{equation}\label{eq:nbartEarth}
\bar{n}_{\mathrm{DM}}\left(t_{\oplus}\right)=\frac{3 f_{\text {cap }} \rho_{\mathrm{DM}} v_{\oplus} t_{\oplus}}{4 m_{\mathrm{DM}} R_{\oplus}} \times \frac{1-\exp \left(-f_{\text {loss }} t_{\oplus}\right)}{f_{\text {loss }} t_{\oplus}}.
\end{equation}

The proper modelling of $f_{\rm{loss}}$ is the crux of determining the final DM number density and possible direct detection signals. NFM18 gives two escape mechanisms that contribute to $f_{\rm{loss}}$: Jeans Escape from the Last Scattering Surface (LSS) and thermospheric Escape from above the LSS.  In the Jeans escape mechanism, captured DM particles thermalize with local Earth's material (crust or gas) and acquire a Maxwell-Boltzmann velocity distribution.  The LSS is defined such that  particles 
with $v>v_{\rm{es}}=11.2 \rm{\,\,km/s}$ and moving upwards will typically escape and not scatter. I.e., the LSS is the surface such that the optical depth above that surface equals unity.   The location of the LSS,  $z_{\rm{LSS}}$, depends on the scattering cross section of DM with Earth's nuclei and is defined through
\begin{equation}
\int_{z_{\rm{LSS}}}^{\infty} \sum_{\mathrm{A}} n_{\mathrm{A}}(z) \sigma_{11}^{\mathrm{A}} d z=1,
\end{equation}
where $\sigma_{11}^{\mathrm{A}}$ is the cross section measured at $11.2 \rm{\,\,km/s}$.  The sum over $A$ includes all nuclei in the atmosphere (z>0) or the crust (z<0).  For the cross sections which can be probed here ($10^{-30} \sim 10^{-20} \mathrm{\,\,cm^2}$), the LSS is always at altitude $z \lesssim 100$ km where the temperature is relatively stable and insensitive to the solar activity compared to the thermosphere.  The fractional loss rate due to Jeans escape is,
\begin{equation}\label{eq:JeansLoss}
f_{\mathrm{loss}}^\text {Jeans}=\frac{3 n_{\mathrm{LSS}} v_{\mathrm{LSS}}}{2 \pi^{1 / 2} R_{\oplus}  \bar{n}_{\mathrm{DM}} }\left(1+\frac{v_{\mathrm{es}}^{2}}{v_{\mathrm{LSS}}^{2}}\right) \exp \left(-v_{\mathrm{es}}^{2} / v_{\mathrm{LSS}}^{2}\right),
\end{equation}
where $n_{LSS}$,  $T_{LSS}$, $v_{\mathrm{LSS}}=\left(2 k T_{\mathrm{LSS}} / m_{\mathrm{DM}}\right)^{1 / 2}$ are the DM number density,  temperature, and velocity, each evaluated at the LSS. 

On the other hand,  above the LSS and $z\gtrsim$ 100 km,  the atmosphere's temperature rises rapidly with $z$.  These hot gas molecules carry much higher energy and can contribute to the loss rate of DM particles by scattering with them.  NFM18 gives the loss rate per unit volume at height z to be
\begin{equation}\label{Lz}
L(z)= \sum_{\mathrm{A}} 2 \pi^{-1 / 2} n_{\mathrm{DM}} n_{\mathrm{A}} \sigma_{11}^{\mathrm{A}} v_{\mathrm{T}} \exp \left(-\frac{1}{2} m_{\mathrm{DM}} v_{\mathrm{es}}^{2} / k T_{\mathrm{eff}}\right)\beta(\tau_z),
\end{equation}
where $T_{\mathrm{eff}}=2 \bar{f}_{\mathrm{KE}} T(z)$. $\beta(\tau)$ is the probability that a particle with $v>v_{\rm{es}}$ actually escapes rather than scattering again, and is given by $\beta(\tau)=\frac{1}{2} \int_{0}^{1} \exp (-\tau / \mu) d \mu $.  $\tau_z$ is the optical depth at $z$ and is determined also by the cross sections:  
\begin{equation}
\tau_z=\int_{z}^{\infty} \sum_{\mathrm{A}} n_{\mathrm{A}}(z') \sigma_{11}^{\mathrm{A}} d z'.
\end{equation}
The fractional loss rate due to thermospheric escape (TE) is
\begin{equation}
f_{\text {loss}}^\text{TE}=\frac{3 \int_{0}^{\infty} L(z) d z }{R_{\oplus} \bar{n}_{\mathrm{DM}}}.
\end{equation}
$f_{\text {loss}}^\text{TE}$ is exponentially sensitive to the temperature at the thermosphere,  which depends strongly on solar activity and could exhibit significant variation over geological timescales.  Due to the lack of solar observation data,  the global averaged $f_{\text {loss}}^\text{TE}$ over the solar circle 1976–1985 is used by NFM18 to estimate $n_{X}$. The limitation of this approximation is beyond the scope of this paper. 
Finally,  the total fractional loss rate is
\begin{equation}
\label{eq:floss}
f_{\rm{loss}} = f_{\mathrm{loss}}^\text {Jeans} + f_{\text {loss}}^\text{TE}.
\end{equation}
Given $m_{X}$ and $\sigma_{11}^{A}$ for all $A$ in the Earth, which depend on specific underlying DM model and parameters,  Eq.~\eqref{eq:nbartEarth} together with Eq.~\eqref{eq:profileDIff} can be used to obtain the DM density profile captured by Earth, $n_{X}(r)$ and in particular the surface density $n_{\mathrm{DM}}\left(R_{\oplus}\right)$ which is relevant for direct detection experiments and the dewar experiment described below.

\subsection{\label{sec:NB} Dewar Experiment}
NFM18 proposed a number of ways to exploit the large density of thermalized DM at the surface of the Earth, to place new limits on the cross section for DM-nucleus scattering.  These include drag on the Hubble Space Telescope, limit on the lifetime of the LHC proton beams due to scattering on DM in the beam pipe, and elevated evaporation rate of cryogens due to DM penetrating the dewar walls and scattering on the cryogen.  The drawback of the limits of NFM18 is that they are for abundance-weighted cross sections over various nuclei and cannot be easily interpreted given that Born approximation is not generally applicable.  To address this, Neufeld and Brach-Neufeld~(NBN19) performed a  series of dedicated dewar experiments to place limits individually on the DM scattering cross section on 27 different nuclei. 

NBN19 immersed samples of the key nuclei composing the Earth's atmosphere and crust, in liquid nitrogen storage dewars.  If a sample is heated by scattering with the thermalized DM particles, that heat will be transfered to the LN and increase its evaporation rate.  Null results are used to place constraints on DM-baryon scattering cross sections.  Given the DM mass $m_{X}$ the quantity being constrained is $n_{\mathrm{DM}}\left(R_{\oplus}\right)\sigma_{\rm{300K}}^{A}$,  which determines the heating rate for the sample.  $\sigma_{300K}^{A}$ stands for the average momentum transfer cross section between DM and nucleus A,  at room temperature $T=300\rm{K}$ with mean speed $\bar{v}=\left(8 k T_{\mathrm{DM}} /\left(\pi m_{\mathrm{DM}}\right)\right)^{1 / 2}$. Notice that for a velocity dependent cross section,  $\sigma_{\rm{300K}}^{A}$ can be different from the cross section for scattering at escape velocity which enters $\sigma_{\rm{11}}^{A}$.
 
\section{\label{sec:Yukawa} Constraints on Yukawa Interaction from the NBN19 dewar experiment}
Obtaining limits on $\sigma_{Xp}$ from the NBN19 limits on heating, is quite involved.  The general approach is, in brief, the following; details are explained in individual subsections which follow. Given any model of DM candidate with mass $m_{X}$,  its scattering cross section with baryons can be expressed as a function of the underlying parameters of the interaction, which we aim to constrain.  (For a Yukawa interaction the parameters are the Yukawa coupling strength $\alpha$, the mediator mass $m_\phi$, and the nuclear mass distribution.)  For a given choice of parameters, we calculate $\sigma_{11}^{A}$.  This, together with $m_{X}$ fixes $n_{\mathrm{DM}}\left(R_{\oplus}\right)$ as described in section~\ref{sec:NFM}.  For the same parameters we calculate $\sigma_{300\rm{K}}^{A}$ and check if $n_{\mathrm{DM}}\left(R_{\oplus}\right)\sigma_{300\rm{K}}^{A}$ is excluded by the dewar experiment results described in section~\ref{sec:NB} or not.  In this way, each point in the ($m_{X},\alpha$) parameter space can be excluded or is allowed.

For WIMPs,  Born approximation is applicable and one can assume the simple scaling relationship between DM-nucleus (with atomic mass number A) cross section $\sigma_{A}$ and DM-nucleon cross section $\sigma_p$,
\begin{equation}
\label{eq:BornScaling}
\sigma_{A}^{\text {Born }}=\sigma_{p}\left(\frac{\mu_{A}}{\mu_{p}}\right)^{2} A^{2}.
\end{equation}
As a result, different direct detection experiments with different target nucleus can put limits on $(m_X, \sigma_p)$ and be compared in a model-independent manner.  However, it has been pointed out~\cite{Xu:2019, Digman:2019,Farrar:2020,Xu:2020qjk_v2} that when the cross section is large and the interaction is non-perturbative,  the cross section is a highly non-trivial function of the mass number $A$ and the constraints obtained from experiments cannot be translated to $\sigma_p$ according to Eq.~\eqref{eq:BornScaling}.  This is the case for the DM captured by Earth and the probed by the dewar experiment. 

\subsection{Interaction between DM and nucleus}

In this paper we adopt the generic model of DM studied in Ref.~\cite{Xu:2020qjk_v2} (XF21 below),  where scalar DM particles with mass $m_{X}$ interact with a nucleon through the exchange of a scalar mediator of mass $m_\phi$.  A Yukawa potential is the most general potential between pointlike particles:
\begin{equation}
\label{eq:yukawa}
V(r)=-\frac{\alpha}{r}e^{-m_\phi r},
\end{equation}
with $\alpha > 0 $ for an attractive interaction.  (For repulsive interaction we replace $\alpha \rightarrow -\alpha$ to keep alpha positive in the plots.) The range of the interaction described by Eq.~\eqref{eq:yukawa} is $m_\phi^{-1}$, and $m_\phi \rightarrow 0$ results in a Coulomb interaction; for sexaquark DM the mediator $\phi$ is the flavor-singlet combination of $\omega$ and $\phi$ vector mesons with mass about 1 GeV.

\begin{figure}
\centering 
\includegraphics[width=.7\textwidth]{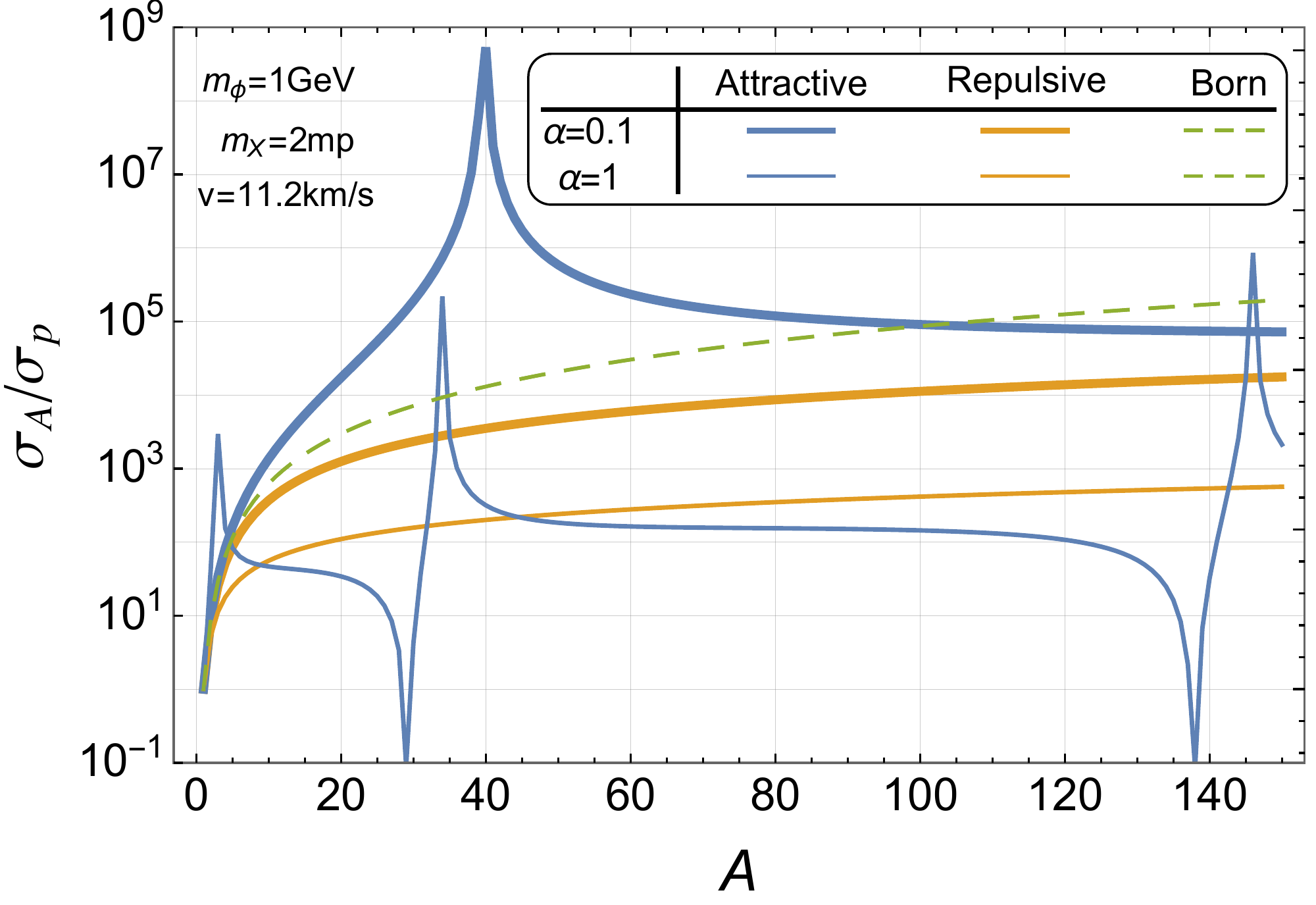}
\caption{\label{fig:Ascaling} $\sigma_A/\sigma_p$ as a function of $A$.  Blue and tan lines are calculated numerically for attractive and repulsive potential~\eqref{eq:Vball}, with $r_0=A^{\frac{1}{3}} \rm{\,\, fm}$.  Green dashed line is the prediction from Born approximation~\eqref{eq:BornScaling}.}
\end{figure}

As in XF21, the finite size of the nucleon and nucleus is accounted for by integrating Eq.~\eqref{eq:yukawa} within a uniform spherical charge density of radius $r_0$, 
\begin{equation}
\label{eq:Vball}
V(r)=-\frac{3 \alpha}{m_{\phi}^{2} r_0^3} \times
\begin{cases}
1-(1+m_\phi r_0) e^{-m_\phi r_0}\frac{\sinh{(m_\phi r)}}{m_\phi r}&\!\!\!\!\! \quad (r < r_0)
\\
\left[ m_\phi r_0 \cosh{(m_\phi r_0)} - \sinh{(m_\phi r_0)}  \right] \frac{e^{-m_{\phi} r}}{m_\phi r}&\!\!\!\!\! \quad (r\geq r_0)~.
\end{cases}
\end{equation}
(As shown in XF21, use of a form factor is not in general a correct procedure to represent an extended source in the non-perturbative regime.)
For nucleus with atomic number A,   we take $\alpha \rightarrow A\alpha$.  $r_0 = R_0 A^{\frac{1}{3}} $ is the nucleus radius and $R_0 \sim 1$ fm.\footnote{ We have checked the sensitivity of our results to the uncertainty in the nuclear wave function by taking different $R_0$ values  (0.9-1.2 fm) and using a different source distribution (uniform sphere vs. Wood-Saxon potential).  This basically just moves the exact location of resonances, e.g., as seen in Fig.~\ref{fig:Ascaling}. Thus except for the details of the ``frilly edges" seen in e.g. Fig.~\ref{fig:attractiveU238He4} and Fig.~\ref{fig:sigmapNFM}, our results are insensitive to the assumptions of the potential and we report values for the potential Eq.~\eqref{eq:Vball} with $R_0 = 1$ fm.}  As described in XF21,  when $\alpha \sim \mathcal{O} (0.1)$ or larger,  the interaction is so strong that Born approximation is no longer accurate.  One must solve the Schrödinger equation numerically to obtain the phase shift and the cross section.  

There are some general features in this non-perturbative regime.  (1) The cross section is s-wave dominant due to the low energy involved,  so the scattering is isotropic and we do not distinguish between the total cross section and the momentum transfer cross section (2) When the interaction is attractive,  there are resonances and anti-resonances where the cross section gets enhanced or reduced by several orders of magnitude.  The resonance corresponds the emergence of a zero energy bound state in the potential~\eqref{eq:Vball}. (3) The cross section is velocity independent at low velocities, except on an attractive resonance where it behaves like $\sigma \sim v^{-2}$; near to but not on a resonance, the velocity dependence is complicated, with both constant and $v^{-2}$ regimes  -- see XF21 for details.

With the model interaction specified, we place constraints on the Yukawa parameter space ($\alpha$, $m_X$) for $m_\phi = 1$ GeV by calculating the cross sections for relevant $A$'s and velocities, which using the NFM18 analysis gives the local number density at the surface of the Earth, then using result of the dewar experiment we decide if the a given value of ($\alpha$,  $m_X$) is allowed or excluded. We pick $m_\phi=1$ GeV as a benchmark to show the non-triviality of the analysis.  

Figure~\ref{fig:Ascaling} shows $\sigma_A / \sigma_p$ as a function of $A$ for $A \leq 150$ for $m_\phi=1$ GeV. The deviation of the $A$ scaling from Born approximation~\eqref{eq:BornScaling} is evident, especially for the attractive interaction where different nuclei can be close to resonance or anti-resonance. The repulsive interaction doe not have resonances and the scaling with $A$ is much weaker than predicted by Born approximation.

\subsection{The fractional loss rate $f_{\rm{loss}}$}
The sensitivity of the dewar experiment depends on the density of the thermalized DM on the surface of Earth, which in turn is determined by the fractional loss rate $f_{\rm{loss}}$ defined in Eq.~\eqref{eq:floss}. As noted, there are two contributions to the loss rate, Jeans escape and thermospheric escape (TE).   
To calculate $f_{\rm{loss}}(\alpha,\,m_X)$ we must separately evaluate the contributions of Jeans and thermospheric escape.  Fig. 5 of NFM18 gives $f_{\rm{loss}}^{\rm{Jeans}} (\sigma_{11}^{\rm es})$, where $\sigma_{11}^{\rm es}$ is the weighted sum of $\sigma_{11}$'s for different nuclei, $A$, appropriate to atmosphere or crust, depending on where the Last Scattering Surface is. For $\sigma_{11}^{\rm atm} \gtrsim 2.5 \times 10^{-26} {\rm cm}^2$ the Last Scattering Surface is in the atmosphere; otherwise it is in the crust.  Thus for each $(\alpha,\,m_X)$ we calculate $\sigma_{11}^{\rm atm}$ and if it is $ \lesssim 2.5 \times 10^{-26} {\rm cm}^2$, we set $\sigma_{11}^{\rm es} = \sigma_{11}^{\rm atm}$.  If not, we set $\sigma_{11}^{\rm es} = \sigma_{11}^{\rm cr}$.  
The thermospheric escape $f_{\rm{loss}}^{\rm{TE}}(\sigma_{11}^{\rm atm})$ can be calculated by the weighted sum of the contributions of individual atmospheric nuclei $A$, given in Fig. 16 of NFM18. 
The sophistication of the NFM18 calculation of the DM atmosphere of Earth, including temperature and density profiles for dozens of nuclei, are embedded within these NFM18 results for $f_{\rm{loss}}^{\rm{Jeans}} (\sigma_{11}^{\rm es})$ and $f_{\rm{loss}}^{\rm{TE}}(\sigma_{11}^{\rm atm})$. We thank the authors for providing tabular values of the results displayed in their Figs. 5 and 16. 


\begin{figure}
\centering 
\includegraphics[width=.7\textwidth]{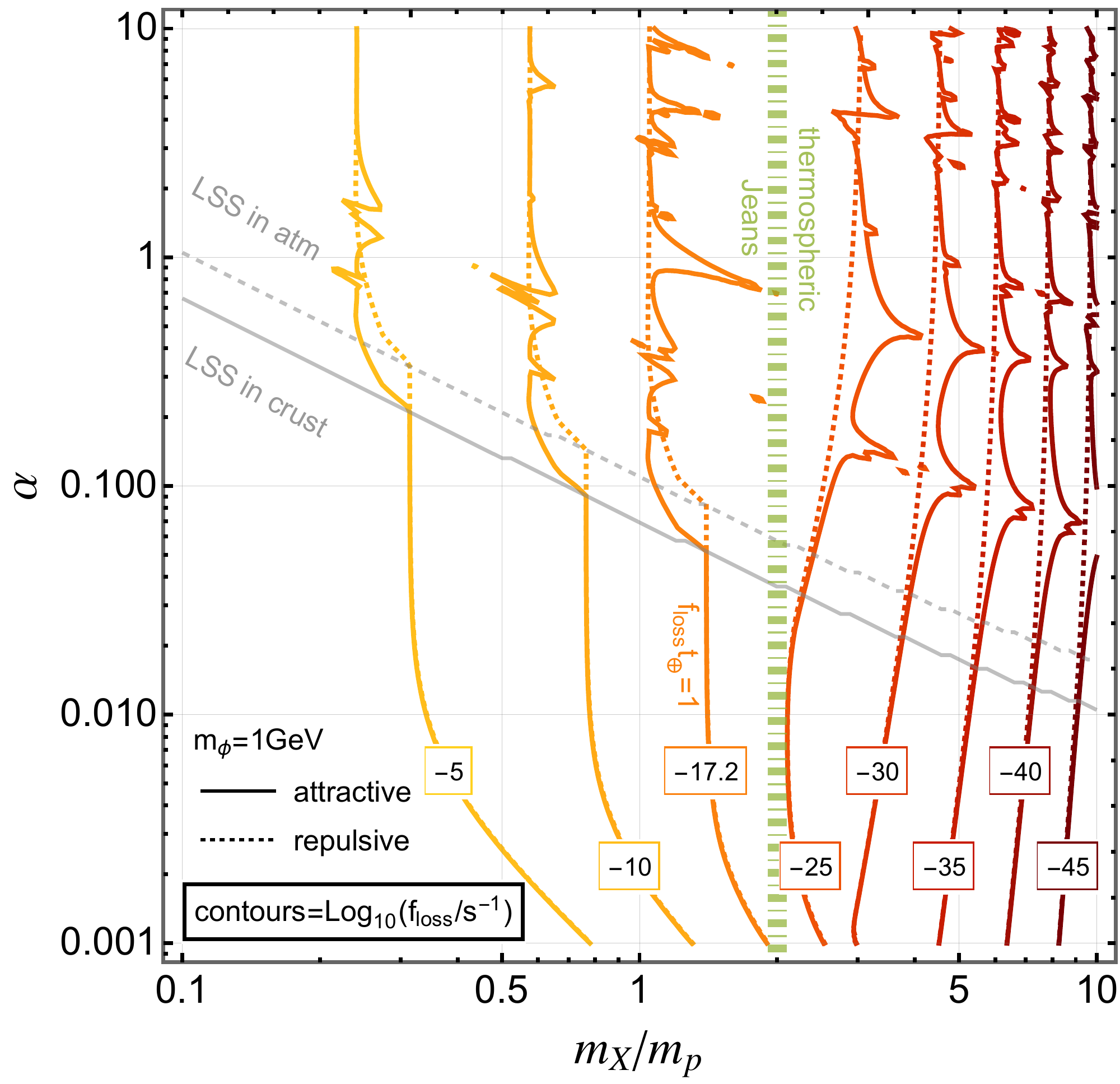}
\caption{\label{fig:floss}$f_{\rm{loss}}(\alpha,m_X)$ for $m_\phi = 1$ GeV, with solid and dashed lines for attractive and repulsive interaction respectively. Darker color corresponds to smaller $f_{\rm{loss}}$ and the numbers on the contour lines indicate $\rm{Log}_{10}(f_{\rm{loss}}/s^{-1})$. The "-17.2" contour is where $f_{\rm{loss}} t_{\oplus} = 1$.  The gray solid (attractive) and dashed (repulsive) line indicate where the LSS drops into the Earth's crust from the atmosphere.  The thick-thin zone at $m_X = 2$ GeV demarcates where Jeans escape dominates over thermosphreic escape or vice versa.}
\end{figure}

Figure~\ref{fig:floss} shows the total loss rate function $f_{\rm{loss}}(\alpha,m_X)$ for $m_\phi=1$ GeV.
The Jeans loss and thermospheric loss can be ignored only when $f_{\rm{loss}} t_{\oplus} \ll 1$, i.e., to the right of the $\rm{Log}_{10}(f_{\rm{loss}}/s^{-1})=-17.2$ contour line or roughly $m_X \gtrsim  2 \, m_p$ in Fig.~\ref{fig:floss}.  On the other hand, if $f_{\rm{loss}} t_{\oplus}$ is too large,  captured $n_{X}$ will be too small to give any interesting new signals.  
In general, as $m_X$ increases,  the Maxwell-Boltzmann distribution will concentrate to the low speed side as $v_{\rm{rms}}=\sqrt{3kT/m_X}$ decreases, reducing the probability of $v>v_{\rm{es}}$ and hence $f_{\rm{loss}}$.  
The gray solid (attractive) and dashed (repulsive) lines shows where the LSS moves into the Earth's crust from the atmosphere as $\alpha$ and $\sigma_A$ decrease.  This transition leads to a discontinuity in $f_{\rm{loss}}$ as can be seen from the contours $\rm{Log}_{10}(f_{\rm{loss}}/s^{-1})=(-17.2,-10,-5)$, where Jeans escape dominates and LSS matters.
Roughly speaking, for $m_X \gtrsim 2$ GeV or to the RHS of the thick-thin green line, thermospheric escape is much larger than Jeans escape and the position of LSS is not important, so there is no discontinuity in $f_{\rm{loss}}$.
The gray lines for the LSS transition lie close to where Born approximation breaks down. A large cross section, which cannot be achieved by Born approximation, is needed for scattering to happen in the relatively thin atmosphere.  When the LSS is in the crust, we are largely in the regime of applicability of Born approximation and attractive and repulsive interactions are indistinguishable. Above the LSS lines, attractive and repulsive interaction give different prediction and the non-perturbative effects of attractive resonance produce the jagged behavior seen in the figure.


\subsection{\label{sec:nDM}The number density at surface $n_{X}(R_{\oplus})$}

\begin{figure}
\centering 
\includegraphics[width=1.0\textwidth]{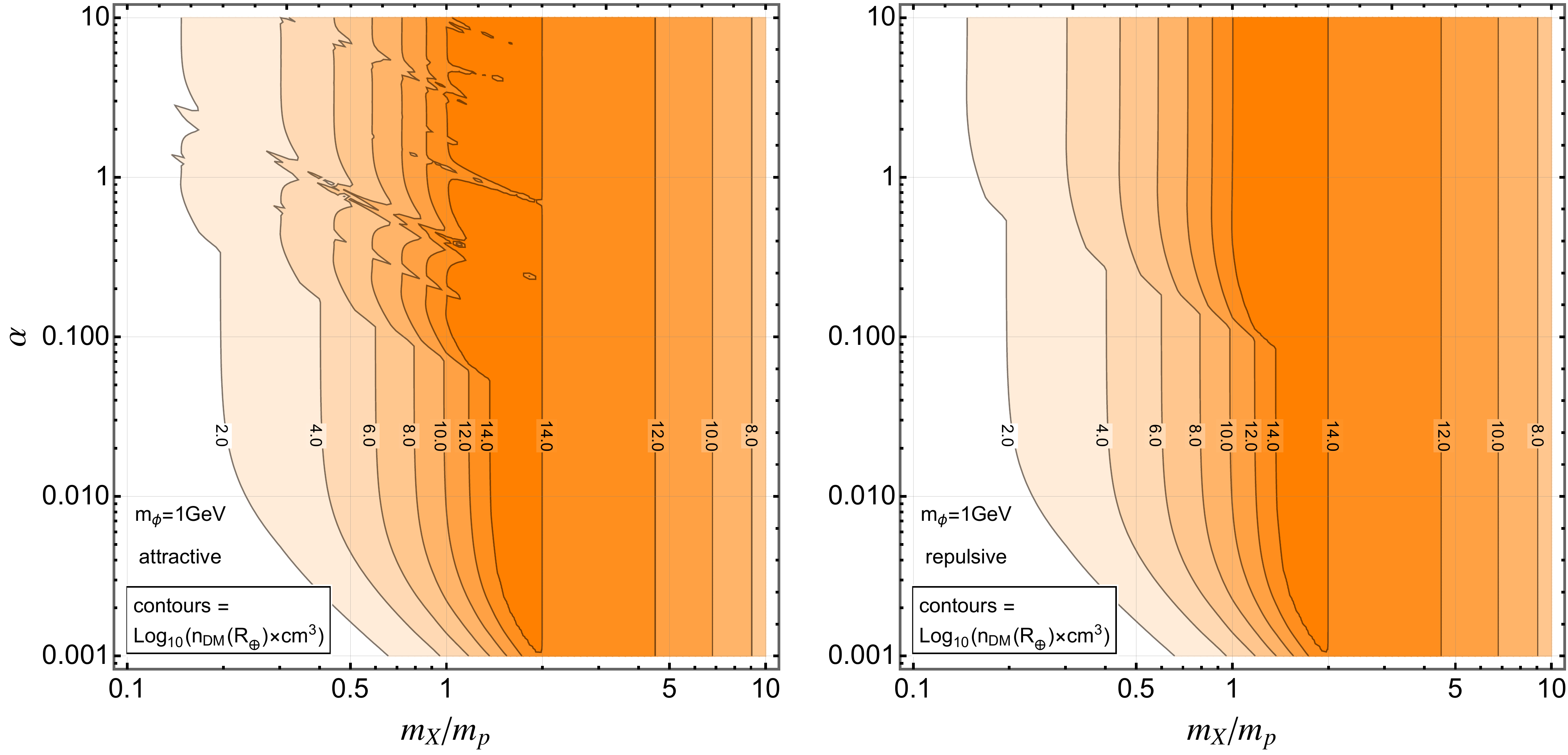}
\caption{\label{fig:nsurface} The number density of captured DM at Earth's surface for attractive (left) and repulsive (right) interaction, calculated assuming $m_\phi=1$ GeV}
\end{figure}

With the knowledge of $f_{\rm{loss}}$ and the DM profile $n_{X} (r)$ obtained by solving Eq.~\eqref{eq:profileDIff}, we can calculate the density of captured DM at Earth's surface. Figure~\ref{fig:nsurface} shows  $n_{X}(R_{\oplus})$ in the ($\alpha,m_X$) plane with $m_\phi=1$ GeV.  
For $m_X \gtrsim$ 2 $m_p$,  the loss rate of DM can be ignored with $f_{\rm{loss}} t_{\oplus} \ll 1$. As a result,  the DM number density is only a function of DM mass $m_X$ and is independent of the cross sections and $\alpha$,  leading to a vertical contour. As $m_X$ continues to increase,  NFM18 showed that the surface DM density drops rapidly because the DM density profile becomes more and more concentrated towards the center of the Earth.
For $m_X \lesssim 2$ $m_p$,  the surface density of DM drops rapidly as $m_X$ decreases due to the increasing of $f_{\rm{loss}}$.  In the perturbative regime for $\alpha \lesssim 0.1$ and when Jeans escape is significant for $m_X \lesssim$ 2 $m_p$, the density also drops as $\alpha$ decreases (as does the cross section) because the LSS enters deeper into the crust where the temperature and $f_{\rm{loss}}$ are higher.
The  contours for a repulsive interaction are similar to those for the attractive interaction, without the resonant structures at $\alpha \gtrsim 0.1$.

From Fig.~\ref{fig:nsurface} we can see for $m_X\! \sim \!(1-2) \, m_p$, the captured DM at the Earth's surface can have $\sim 14$ orders of magnitude more concentration than that of the Galactic average.  These thermalized DM have energy $E \sim 3/2 \times 300\rm{K} \sim 0.04$ eV and small velocities $ <v> = \sqrt{3 T/m} = 2.64 \rm{\,\, km/s} \,\, (m_X/{\rm GeV})^{-1/2} \ll  v_{\rm esc} = 11.2$ km/s, producing an energy deposit way below the threshold energy of existing direct detection experiments.  Nonetheless, due to their high concentration they can have detectable signals in the dewar experiment if they are not bound to nuclei.

Our Fig.~\ref{fig:nsurface} for $n_{X}(R_{\oplus})$ as a function of $(\alpha,m_X)$ can be compared to Fig. 6 of NFM18 where $n_{X}(R_{\oplus})$ is shown as a function of $(\sigma_{11}^{\rm{es}}, m_X)$, where $\sigma^{\rm{es}}$ is a weighted sum of $\sigma_{11}^A$. Both figures share some similar features since $\sigma_{11}^{\rm{es}}$ is roughly a increasing function of $\alpha$ except near (anti-)resonances in the case of an attractive interaction.

\subsection{Exclusion on $(\alpha,m_X)$}
With $n_{X}(\alpha,m_X)$ and $\sigma^A_{\rm{300K}} (\alpha,m_X)$ we can place constraints on the $(\alpha,m_X)$ plane from the dewar experiment, for each nucleus $A$.  Notice that $\sigma^A_{\rm{300K}}$ is relevant for the heating of samples in the dewar while $\sigma^A_{\rm{11}}$ is used to determine $n_{X}$ in the DM atmosphere.  These two cross sections are evaluated at two different characteristic velocities for each nucleus.  In most of the parameter space they are the same except on or near resonance where $\sigma_A \sim v^{-2}$.  Calculating them separately gives us more accuracy near the resonance. See XF21 for more information on velocity dependence.

\begin{figure}
\centering 
\includegraphics[width=1.0\textwidth]{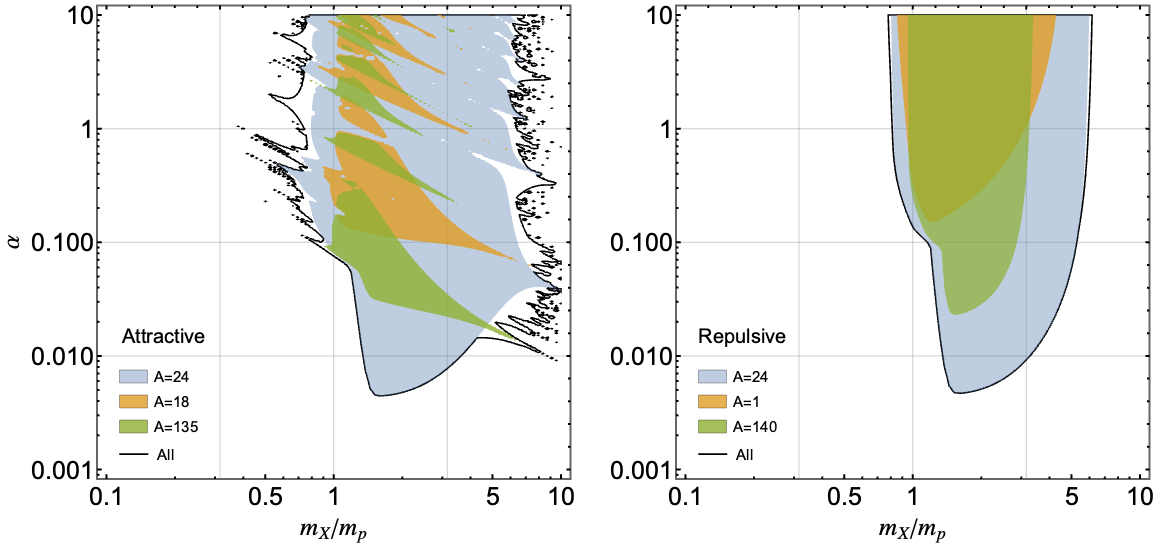}
\caption{\label{fig:alphaNFM} Exclusion region on $(\alpha,m_X)$ with $m_\phi=1$ GeV,  for an attractive (left) or repulsive (right) interaction, applicable if the DM does not bind to nuclei. Colored regions shows limits from certain nuclei. Black contours show limits for all nuclei combined.  $A=$24 has a large contribution in both cases.}
\end{figure}

Figure~\ref{fig:alphaNFM} shows the exclusion region on $(\alpha,m_X)$ for 3 selected nuclei from the dewar experiment (colored region) as well as the exclusion region for 74 nuclei combined (black contour).  Due to the non-perturbative scaling of cross sections as a function of $\alpha$, $m_X$ and $A$, and the variations of the experimental limit for different nuclei, different $A$ cover different region in the parameter space and one nucleus can be complimentary to another. Figure~\ref{fig:alphaNFM} gives a example of this complimentary coverage for the attractive ($A=$18 and 135) and repulsive ($A=$1 and 140) interaction.  This effect is especially important for the attractive interaction, where the exclusion region is determined by the location of resonances and anti-resonances, which can be quite different for different nuclei as shown in Fig.~\ref{fig:Ascaling}. 

The jagged peninsula-like structures at the boundary of the attractive figure are due to the (anti-)resonances.  As emphasized in XF21, the exact location of the resonances are sensitive to $m_\phi$ and the nuclear wave function which we modelled crudely as a finite square well.  So the boundary should be considered blurry.  Conservatively, we just remove the peninsulas in final bounds.

As a result, in order to thoroughly explore the parameter space,  a scan over atomic mass number $A$ is important. This was done in the dewar experiment by changing the sample in the dewar and doing the measurement again. Combining results from experiments with different target nuclei also helps to close the gaps, as was done previously in XF21 for XQC ($A=$28), CMB ($A=$1,4) and CRESST (various $A$ involved).

The exclusion regions for each individual nuclei are included in the appendix (Fig.~\ref{fig:alphaAllAttractive}~and~\ref{fig:alphaAllRepulsive}),  where the contribution from each $A$ is shown.  For both the attractive and repulsive interaction, the largest contribution to the overall exclusion happens to come from Mg-24,  which gives $\sigma_{\rm{300K}}^{24} \leq 10^{-27} \rm{\,\,cm^2}$ according to NBN19 (for $n_{X} = 10^{14} \rm{\,\,cm^3}$ and $m_X=2m_p$).  In the dewar experiment, Mg-24 generates one of the strongest constraints for all nuclei.  In fact, several nuclei give fairly strong constraints,  including N-14, Mg-24 and Al-27. On the other hand, certain nuclei gave poor limits on the heating rate thus excluding little area in the $(\alpha,m_X)$ plane,  e.g.  Sr-38 and Ba-130.  In the non-perturbative regime, there is no a priori way to know  which nucleus will give the best result.

As seen in Fig.~\ref{fig:alphaNFM}, the dewar experiment is best for constraining $m_X \sim 0.5-10$ GeV, beyond which the DM density is too low according to Fig.~\ref{fig:nsurface}, requiring a cross section $\sigma_{\rm{300K}}^A$ needed to saturate the limit that cannot be reached for any $(\alpha,m_X)$ in our Yukawa potential model. Our work here, assuming pointlike DM interacting through a Yukawa potential, could be elaborated to apply to DM with an extended spatial distribution, allowing limits to be placed on the size of DM particles from the dewar experiments.

\subsection{Possibility of DM capture by nuclei}
\label{sec:capture}
A caveat concerning the analysis of the previous section is that -- in the case of an attractive interaction -- capture of a DM particle by a nucleus removes that DM particle from participation in heating the cryogens.  If an attractive potential is strong enough to generate a resonance, it is also capable of accommodating DM-nucleus bound states.  In fact, the resonance in the elastic scattering cross section exactly corresponds to a zero energy bound state, a familiar result from introductory quantum mechanics~\cite{Sakurai:QM}.  Larger values of $\alpha$ imply deeper and deeper binding, eventually leading to multiple bound states for a given nucleus. 

\begin{figure}
\centering 
\includegraphics[width=0.7\textwidth]{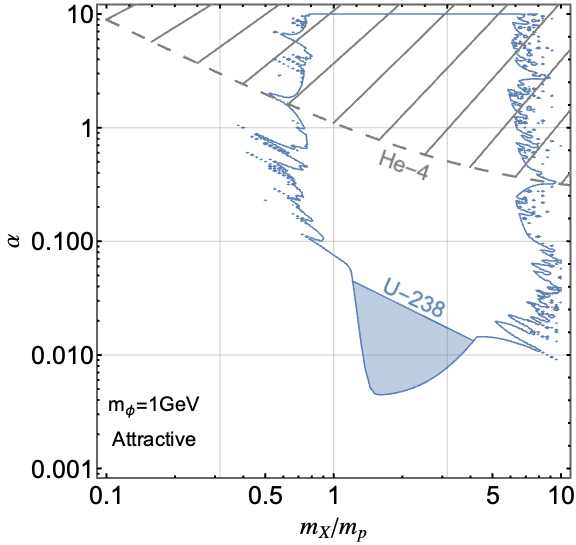}
\caption{\label{fig:attractiveU238He4} Exclusion region on $(\alpha,m_X)$ for $m_\phi=1$ GeV for an attractive interaction, taking into account DM-nucleus binding.  The cross-hatched region is excluded by requiring DM and $^4$He cannot bind.  If DM can bind with U-238 or lighter nuclei, the dewar experiment loses sensitivity so the white region above the U-238 line cannot be excluded based on bounds from the dewar experiments. Everything within the blue contour is conservatively excluded, assuming a full DM atmosphere in the absence of binding for $A>238$. With additional work on the constraints on DM binding, the upper edge of the exclusion region can likely be pushed up marginally.}
\end{figure}

An important feature of the extended Yukawa potential Eq.~\eqref{eq:Vball}, is that for a fixed $\alpha$, there is a minimum $A$ for the formation of a bound state.  $A_{\rm{min}}(\alpha)$ is a decreasing function of $\alpha$, so for $\alpha$ large enough essentially all nuclei have bound states with DM particles.  This leads to an immediate requirement, that $\alpha$ must be small enough that DM does not bind to $^4$He (cross-hatched region in Fig.~\ref{fig:attractiveU238He4}), otherwise a large fraction of the $^4$He produced in primordial nucleosynthesis would actually be a $^4$He-DM bound state, and helium atoms would have a mass roughly $m_X$ larger than observed. 

The DM capture process is similar to neutron capture by a heavy nucleus. A detailed discussion of the capture process, including calculation of capture cross sections and abundances of the exotic bound state nuclei will be presented elsewhere~\cite{Xu:2021cap}.  In the following, we assume -- in a very good approximation -- that all thermalized DM are bound to some nucleus, if $\alpha$ is large enough that a bound state exists for any sufficiently abundant nucleus in the Earth. 

Taking into account that dewar experiments are not constraining if a DM atmosphere does not exist due to DM binding to nuclei, we conservatively identify the $\alpha$ range such that DM does not bind to $^{238}$U, one of the heaviest yet still abundant elements in Earth's crust. In Fig.~\ref{fig:attractiveU238He4}, the white region within the originally excluded range (left panel of~Fig.~\ref{fig:alphaNFM}) shows where thermalized DM binds to U-238 or lighter elements, and the dewar experiment loses its sensitivity.  If Pb rather than U were the
lightest mass nucleus able to bind DM, the reach in $\sigma_p$ increases by $< 20$\%.  

One further potential caveat for using the dewar experiments' limits, is that they rely on the wall of the dewar (made of $^{27}$Al) being transparent to the DM particles.  If instead the DM particles thermalize inside the wall rather than heating the liquid inside the dewar, the derived limits are invalid.  NFM18 obtained a conservative gauge of the dewar technique's validity: that the DM-$^{27}$Al scattering length be longer than the thickness of the dewar, $\sim 1$ cm.  To saturate that limit requires $\sigma_{300K}^{27} \gtrsim 4\times 10^{-21} \rm{\,\, cm^2}$.  (A more detailed calculation would take into account that multiple scatterings are required to reduce the effective temperature of DM from 300K to the liquid nitrogen temperature, $\sim$ 77K, requiring a still larger cross section.)  A cross section this large cannot be achieved with a repulsive interaction and pointlike dark matter, which is bounded by the range of the potential ~\cite{Digman:2019,Xu:2020qjk_v2}. If the interaction is attractive, this value is still very large and almost saturates the s-wave unitary bound $\sigma_{\rm{s-wave}}\leq 4\pi/(\mu v)^2$ where $\mu$ is the reduced mass (see XF21). However $\alpha$ producing a resonance for Al is well-within the regime where the DM is expected to be bound to He-4, which is excluded by BBN, so this caveat does not impact the applicability of dewar experiment limits.

\section{\label{sec:cxlimits}Exclusion on $(\sigma_p,m_X)$ and comparison with other limits} 
We can now go from our exclusion on ($\alpha,m_X$) to the excluded region in the DM-proton cross section and DM mass $(\sigma_p, m_X)$ plane, using the exact relationship for $\sigma_p(m_X,\alpha)$ calculated for the extended Yukawa potential (taking the proton to be a sphere of radius 1 fm).  This enables us to compare the dewar experiment limits with the limits obtained in XF21 from XQC, CMB,  CRESST and Milky Way gas cloud cooling.  

For a repulsive potential the procedure is straightforward because there is a one-to-one correspondence between $\alpha$ and $\sigma_p$ so an excluded ($\alpha,\, m_X$) implies an excluded ($\sigma_p,\, m_X$).  However, as can be seen in Fig. 9 of XF21, there is not a 1-1 correspondence for an attractive interaction in the resonance region.
This generates a complication in deriving the excluded region in  ($\sigma_p,\, m_X$) for an attractive interaction due to the resonances, which we also encountered in XF21. (See Fig.11 and corresponding discussion in XF21.) 

The procedure adopted is the following.  First, focus on the range $0.001 < \alpha \leq 10$.  For a particular $m_X$, a given value of $\sigma_p$ is allowed if there exists an $\alpha \leq 10 $ such that that $\alpha$ is allowed in~Fig.~\ref{fig:alphaNFM} (left, for the attractive case), or in Fig.~\ref{fig:attractiveU238He4} if DM-nucleus binding makes dewar limits inapplicable, and $\sigma_p=\sigma_p(\alpha)$ calculated from the Yukawa potential~\eqref{eq:Vball}. Any $\sigma_p > \sigma_p(\alpha=0.001)$ that is not allowed will be excluded. This minimum value $\alpha=0.001$ is small enough that Born approximation holds and $\sigma_p(\alpha)$ is a monotonic function below this point, so that no smaller $\alpha$ can produce the given $\sigma_p$.  However, because $\sigma_p$ is not monotonically increasing for $\alpha>10$ due to the existence of (anti-)resonances, it is possible that some $\sigma_p$ not possible for smaller $\alpha$ will turn out to be realized for a higher allowed $\alpha>10$. 
While this complication was important for the analysis of XF21a, the issue does not arise when interpreting the dewar experiments, thanks to our having derived an upper limit on $\alpha$ in the attractive case from non-binding of DM and $^4$He, which is within the test range $\alpha<10$ for $m_\phi=1$ GeV and $m_X=0.1-100$ GeV, so determination of the excluded region in ($\sigma_p,\, m_X$) is robust.


\begin{figure}
\centering 
\includegraphics[width=0.7\textwidth]{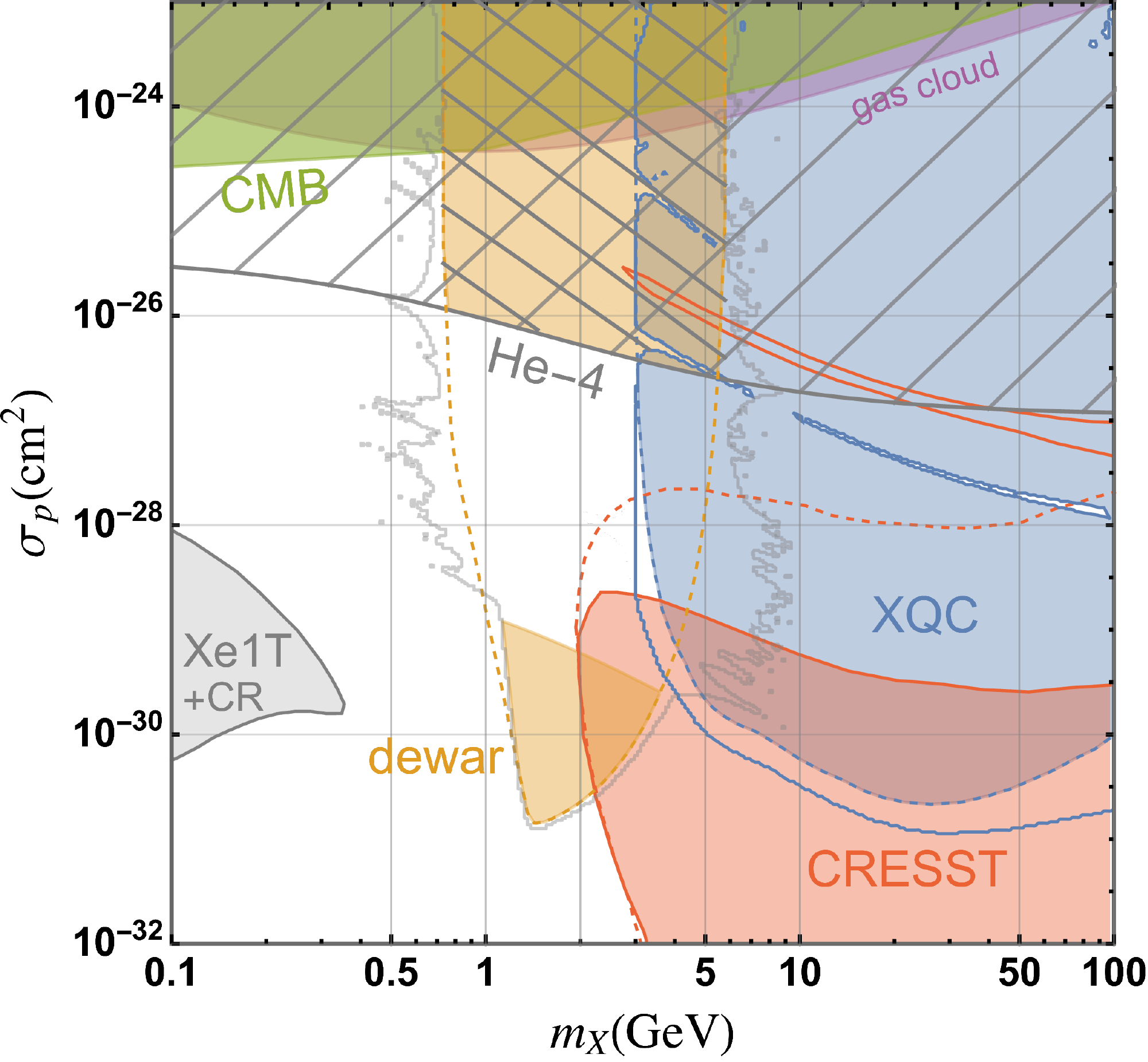}
\caption{\label{fig:sigmapNFM} The new exclusion regions derived here, on ($\sigma_p,m_X$) from the Neufeld dewar experiment are shown in orange.  Excluding the upper cross-hatched portion is only possible thanks to having both the dewar limit -- applicable for repulsive interactions -- and the non-existence of exotic $^4$He-DM, in the attractive case.
Previous constraints obtained in XF21 and~\cite{Bringmann:2018cvk} are also shown.  Colored regions are conservatively excluded. Solid and dashed lines show the perimeter of additional regions which become excluded if the sign of the interaction is specified to be attractive (solid) or repulsive (dashed). The window in the dewar experiment for an attractive interaction is above the binding of DM to U-238.   The light grey contour indicates the boundary of the region (imprecise in its detail) which could be excluded if the interaction were attractive but the dewar limit were nevertheless applicable; a concrete realization of such a scenario is not known. The plot is valid for mediator mass $\geq 1$ GeV; for light mediators it is only approximate as discussed in the text.}
\end{figure}

Figure~\ref{fig:sigmapNFM} shows the exclusion region in the ($\sigma_p,m_X$) plane for $m_\phi=1$ GeV, for both attractive and repulsive interactions.  The conservative limits obtained in XF21 from XQC, CMB, CRESST and Milky Way gas clouds using a fully non-perturbative treatment rather than the previously-used incorrect Born-approximation scaling of cross section dependence on $A$, with similar procedures as the present paper, and the Xe1T limits at low mass using \cite{}, are also shown.  Colored regions are robustly excluded, whether the interaction is attractive or repulsive, while the solid and dashed lines mark the boundaries of additional excluded parameter space if the sign of the interaction is determined or postulated to be attractive (solid) or repulsive (dashed).  The orange region with cross-hatching is excluded for an attractive interaction by non-observation of exotic $^4$He-DM bound states produced in BBN, and by the dewar experiment if the interaction is repulsive.  Additionally, in case the interaction is attractive but for some reason DM binding to nuclei does not evade the dewar limits, the limit which would be derived from the dewar experiment is indicated by the light grey contour, with the exact positions of the jagged ``peninsulae" dependent on uncertain details of the nuclear wave function.

Our results are displayed in Fig.~\ref{fig:sigmapNFM} and earlier, for a mediator mass $m_\phi=1$ GeV.  However these plots apply rather accurately to $m_\phi >1$ GeV as well.  This is because when $m_\phi \gtrsim $ GeV (so that $r_A \gg 1/m_\phi$), the extended Yukawa potential can be approximated by a uniform spherical well with radius $r_A$ and depth $V_{0} \propto \alpha / m_{\phi}^{2}$. As a result, the cross sections $\sigma_A$ are only a function of $\alpha / m_{\phi}^{2}$.  Figures~\ref{fig:alphaNFM} and~\ref{fig:attractiveU238He4}  remain the same for a larger $m_\phi$ except for a rescaling of the $\alpha$ axis according to $\alpha \sim m_\phi^2$ and our limits on ($\sigma_p,m_x$) in Fig.~\ref{fig:sigmapNFM} will not be changed. For lighter $m_\phi$ a general analytic argument is not available, but the limits are quite insensitive to $m_\phi$ down to $\sim 1$ MeV, as shown in XF21. For MeV $\lesssim m_\phi \lesssim$ GeV, the cross sections cannot be obtained by any simple re-scaling of the $m_\phi=$ GeV result and everything needs to be re-calculated. As $m_\phi$ decreases further, there are analytical expressions for the cross section. For 0.1 MeV $\lesssim m_\phi \lesssim$ MeV, Born approximation for a point Yukawa potential can be used. For $m_\phi \ll $ MeV, the classical fitting functions given in~\cite{Khrapak:2004ieee,Loeb:2010gj} provide a good description.  See XF21 for more details.

To recapitulate this section, we have derived the constraints placed by the NBN19 dewar experiment on the interactions of a GeV-mass dark matter particle with nucleons.  The constraints are a valuable addition to direct detection limits from XQC and CRESST which suffer from an uncertain minimum mass threshold due to the unknown thermalization efficiency $\epsilon_{\rm th}$;  this was originally uncritically assumed to be unity, but in fact could be extremely small~\cite{Mahdawi:2018euy, Xu:2020qjk_v2} (Fig.~\ref{fig:sigmapNFM} shows the limits taking $\epsilon_{\rm th} = $ 1\%.)
If the interaction is attractive and DM-nucleus bound states can be formed, there is still an open window for GeV dark matter with a large cross section in the non-perturbative regime: $\sigma_p \sim 10^{-29} \rm{\,\,to \,\,} 10^{-26} \rm{\,\,cm^2}$.  In this region, almost all thermalized DM would bind to nuclei so the dewar experiment is insensitive.  



\section{\label{sec:summary}Summary}

We have derived new limits on the interaction cross section with nucleons, of dark matter in the previously nearly unconstrained $\sim$1-2 GeV mass range.  We exclude a DM-nucleon cross section $\sigma_p> 10^{-26} - 10^{-26.3} {\,\,\rm cm}^2 $, representing roughly an order-of-magnitude improvement over the previous best limits from the CMB and Milky Way gas clouds.  We also exclude a roughly 1-decade domain centered on $10^{-30} {\,\,\rm cm}^2$

Our limits are obtained from an entirely new approach based on the rate of evaporation of cryogens.  For dark matter mass in this range, and DM-nucleus scattering cross sections above $\sim 10^{-30}{\,\,\rm cm}^2$, Neufeld, Farrar and McKee~\cite{Neufeld:2018slx} showed that dark matter interactions with nuclei in the Earth and its atmosphere produce a significant enhancement in the concentration of dark matter bound to the Earth. NFM18 determined the abundance of dark matter, taking into account Jeans and thermospheric losses, and derived several new limits on the DM-baryon cross section exploiting the concentrated atmosphere of thermalized DM particles.  One of the techniques was based on the evaporation rate of cryogens. Neufeld and Brach-Neufeld ~\cite{Neufeld:2019xes} systematically extended these limits by measuring the evaporation rate of liquid nitrogen in dewars in which a diversity of materials had been immersed.  

It is non-trivial to interpret the results of the dewar experiments as a limit on $\sigma_p$, due to the non-perturbative behavior of the cross sections and lack of a simple scaling with $A$.  We have developed a general and rigorous procedure which we applied to the NBN19 results, assuming a Yukawa interaction sourced by the nucleus XF21.  This enabled us to place the constraints shown in Fig.~\ref{fig:sigmapNFM}, which are valid for a mediator mass $m_\phi \gtrsim 1$ GeV.  For MeV $< m_\phi < 1$ GeV our procedure still works and the constraints on the ($\sigma_p,\,m_\phi$) remain roughly the same, but the analysis should be re-done for maximal precision.  

The dewar experiment can probe smaller dark matter mass than the previously-used direct detection experiments such as XQC and CRESST, thanks to the high concentration of thermalized DM and the absence of an energy threshold.  Having an alternative to XQC and CRESST is also important because interpreting those experiments for low mass DM requires making some assumption about the efficiency with which the (very small) kinetic energy of a recoil nucleus is thermalized. Until this is measured, the low-mass reach of these experiments cannot be trusted~\cite{Mahdawi:2018euy,Xu:2020qjk_v2}.  

If the DM-nucleon interaction is attractive and strong enough that dark matter can be captured by sufficiently massive nuclei in the Earth rather than forming a dark matter atmosphere, the dewar experiments are not sensitive.  
In this case, the DM-nucleus bound states appear as exotic isotopes of  nuclei\cite{Farrar:2020}.  Bounds on the abundance of such exotic isotopes, if sufficiently strong, could themselves put constraints on the DM-nucleon interaction. Furthermore, the photon radiated in the capture process could potentially be detected.  We leave analysis related to the DM-nucleus binding to a future work.

In sum, the dewar experiment in combination with the knowledge that $^4$He does not form bound states with the dark matter, leads to the strongest limit to date for $1< m_X\lesssim 3 $ GeV: $\sigma_p < (10^{-26} - 10^{-26.3}) {\,\,\rm cm}^2 $.  The dewar experiments provide a further excluded range on 1.3-4 GeV dark matter:  the region around $\sigma_p \sim 10^{-30} {\,\,\rm cm}^2$. These limits are not yet strong enough to be restrictive on the possibility that dark matter is composed of sexaquarks because the general limit is at the top of the expected range~\cite{fStableS21inprep} and the conditional limit is not constraining, since the sign of the dark matter coupling to the mediator is unknown.  



\acknowledgments
We thank D.~A.~Neufeld for discussions and for providing the data in~\cite{Neufeld:2018slx,Neufeld:2019xes}, and C.~F.~McKee for suggestions on the manuscript. XX received support from a James Arthur Graduate Fellowship and NSF-PHY-2013199; the research of GRF has been supported by NSF-PHY-2013199.


%
%
%
%
%

\bibliography{xxcbib}

\providecommand{\href}[2]{#2}\begingroup\raggedright\begin{thebibliography}{10}

\bibitem{Aprile:2018}
E.~Aprile, J.~Aalbers, F.~Agostini, M.~Alfonsi, L.~Althueser, F.~Amaro et~al.,
  \emph{Dark matter search results from a one ton-year exposure of {XENON}1t},
  \href{https://doi.org/10.1103/physrevlett.121.111302}{\emph{Physical Review
  Letters} {\bfseries 121} (2018) }.

\bibitem{PandaX-II:2020oim}
{\scshape PandaX-II} collaboration, \emph{{Results of dark matter search using
  the full PandaX-II exposure}},
  \href{https://doi.org/10.1088/1674-1137/abb658}{\emph{Chin. Phys. C}
  {\bfseries 44} (2020) 125001}
  [\href{https://arxiv.org/abs/2007.15469}{{\ttfamily 2007.15469}}].

\bibitem{Starkman:1990nj}
G.D.~Starkman, A.~Gould, R.~Esmailzadeh and S.~Dimopoulos, \emph{{Opening the
  Window on Strongly Interacting Dark Matter}},
  \href{https://doi.org/10.1103/PhysRevD.41.3594}{\emph{Phys. Rev. D}
  {\bfseries 41} (1990) 3594}.

\bibitem{Mahdawi:2017cxz}
M.S.~Mahdawi and G.R.~Farrar, \emph{{Closing the window on $\sim$GeV Dark
  Matter with moderate ($\sim$ $\mu$b) interaction with nucleons}},
  \href{https://doi.org/10.1088/1475-7516/2017/12/004}{\emph{JCAP} {\bfseries
  1712} (2017) 004} [\href{https://arxiv.org/abs/1709.00430}{{\ttfamily
  1709.00430}}].

\bibitem{Mahdawi:2018euy}
M.S.~Mahdawi and G.R.~Farrar, \emph{{Constraints on Dark Matter with a
  moderately large and velocity-dependent DM-nucleon cross-section}},
  \href{https://doi.org/10.1088/1475-7516/2018/10/007}{\emph{JCAP} {\bfseries
  1810} (2018) 007} [\href{https://arxiv.org/abs/1804.03073}{{\ttfamily
  1804.03073}}].

\bibitem{McCammon:2002}
D.~McCammon, R.~Almy, E.e..a.~Apodaca, W.B.~Tiest, W.~Cui, S.~Deiker et~al.,
  \emph{A high spectral resolution observation of the soft x-ray diffuse
  background with thermal detectors},
  \href{https://doi.org/10.1086/341727}{\emph{The Astrophysical Journal}
  {\bfseries 576} (2002) 188}.

\bibitem{Wandelt:2000ad}
B.D.~Wandelt, R.~Dave, G.R.~Farrar, P.C.~McGuire, D.N.~Spergel and
  P.J.~Steinhardt, \emph{{Selfinteracting dark matter}},  in \emph{{4th
  International Symposium on Sources and Detection of Dark Matter in the
  Universe (DM 2000)}}, pp.~263--274, 6, 2000
  [\href{https://arxiv.org/abs/astro-ph/0006344}{{\ttfamily
  astro-ph/0006344}}].

\bibitem{Zaharijas:2004jv}
G.~Zaharijas and G.R.~Farrar, \emph{{A Window in the dark matter exclusion
  limits}}, \href{https://doi.org/10.1103/PhysRevD.72.083502}{\emph{Phys. Rev.
  D} {\bfseries 72} (2005) 083502}
  [\href{https://arxiv.org/abs/astro-ph/0406531}{{\ttfamily
  astro-ph/0406531}}].

\bibitem{Erickcek:2007jv}
A.L.~Erickcek, P.J.~Steinhardt, D.~McCammon and P.C.~McGuire,
  \emph{{Constraints on the Interactions between Dark Matter and Baryons from
  the X-ray Quantum Calorimetry Experiment}},
  \href{https://doi.org/10.1103/PhysRevD.76.042007}{\emph{Phys. Rev.}
  {\bfseries D76} (2007) 042007}
  [\href{https://arxiv.org/abs/0704.0794}{{\ttfamily 0704.0794}}].

\bibitem{Xu:2018efh}
W.L.~Xu, C.~Dvorkin and A.~Chael, \emph{{Probing sub-GeV Dark Matter-Baryon
  Scattering with Cosmological Observables}},
  \href{https://doi.org/10.1103/PhysRevD.97.103530}{\emph{Phys. Rev. D}
  {\bfseries 97} (2018) 103530}
  [\href{https://arxiv.org/abs/1802.06788}{{\ttfamily 1802.06788}}].

\bibitem{Wadekar:2019mpc}
D.~Wadekar and G.R.~Farrar, \emph{{Gas-rich dwarf galaxies as a new probe of
  dark matter interactions with ordinary matter}},
  \href{https://doi.org/10.1103/PhysRevD.103.123028}{\emph{Phys. Rev. D}
  {\bfseries 103} (2021) 123028}
  [\href{https://arxiv.org/abs/1903.12190}{{\ttfamily 1903.12190}}].

\bibitem{Hui:2016ltb}
L.~Hui, J.P.~Ostriker, S.~Tremaine and E.~Witten, \emph{{Ultralight scalars as
  cosmological dark matter}},
  \href{https://doi.org/10.1103/PhysRevD.95.043541}{\emph{Phys. Rev. D}
  {\bfseries 95} (2017) 043541}
  [\href{https://arxiv.org/abs/1610.08297}{{\ttfamily 1610.08297}}].

\bibitem{vandenBosch+Shattering19}
N.~Mandelker, F.C.~van~den Bosch, V.~Springel and F.~van~de Voort,
  \emph{Shattering of cosmic sheets due to thermal instabilities: A formation
  channel for metal-free lyman limit systems},
  \href{https://doi.org/10.3847/2041-8213/ab30cb}{\emph{The Astrophysical
  Journal} {\bfseries 881} (2019) L20}.

\bibitem{DES:2020fxi}
{\scshape DES} collaboration, \emph{{Milky Way Satellite Census. III.
  Constraints on Dark Matter Properties from Observations of Milky Way
  Satellite Galaxies}},
  \href{https://doi.org/10.1103/PhysRevLett.126.091101}{\emph{Phys. Rev. Lett.}
  {\bfseries 126} (2021) 091101}
  [\href{https://arxiv.org/abs/2008.00022}{{\ttfamily 2008.00022}}].

\bibitem{Bringmann:2018cvk}
T.~Bringmann and M.~Pospelov, \emph{{Novel direct detection constraints on
  light dark matter}},
  \href{https://doi.org/10.1103/PhysRevLett.122.171801}{\emph{Phys. Rev. Lett.}
  {\bfseries 122} (2019) 171801}
  [\href{https://arxiv.org/abs/1810.10543}{{\ttfamily 1810.10543}}].

\bibitem{Angloher:2017sxg}
{\scshape CRESST} collaboration, \emph{{Results on MeV-scale dark matter from a
  gram-scale cryogenic calorimeter operated above ground}},
  \href{https://doi.org/10.1140/epjc/s10052-017-5223-9}{\emph{Eur. Phys. J. C}
  {\bfseries 77} (2017) 637}
  [\href{https://arxiv.org/abs/1707.06749}{{\ttfamily 1707.06749}}].

\bibitem{Xu:2020qjk_v2}
X.~Xu and G.R.~Farrar, \emph{{Resonant Scattering between Dark Matter and
  Baryons: Revised Direct Detection and CMB Limits}},
  \href{https://arxiv.org/abs/2101.00142v2}{{\ttfamily 2101.00142v2}}.

\bibitem{Farrar:2020}
G.R.~Farrar, Z.~Wang and X.~Xu, \emph{{Dark Matter Particle in QCD}},
  \href{https://arxiv.org/abs/2007.10378}{{\ttfamily 2007.10378}}.

\bibitem{Neufeld:2018slx}
D.A.~Neufeld, G.R.~Farrar and C.F.~McKee, \emph{{Dark Matter that Interacts
  with Baryons: Density Distribution within the Earth and New Constraints on
  the Interaction Cross-section}},
  \href{https://doi.org/10.3847/1538-4357/aad6a4}{\emph{Astrophys. J.}
  {\bfseries 866} (2018) 111}
  [\href{https://arxiv.org/abs/1805.08794}{{\ttfamily 1805.08794}}].

\bibitem{Neufeld:2019xes}
D.A.~Neufeld and D.J.~Brach-Neufeld, \emph{{Dark Matter that Interacts with
  Baryons: Experimental Limits on the Interaction Cross-section for 27 Atomic
  Nuclei, and Resultant Constraints on the Particle Properties}},
  \href{https://doi.org/10.3847/1538-4357/ab15d5}{\emph{Astrophys. J.}
  {\bfseries 877} (2019) 8} [\href{https://arxiv.org/abs/1904.01590}{{\ttfamily
  1904.01590}}].

\bibitem{Xu:2019}
X.~{Xu} and G.~{Farrar}, \emph{{Interaction Between Dark Matter and Baryons:
  Non-Perturbative Effects, Experimental Bounds, and Solution to the 7Li
  Problem}},  in \emph{APS April Meeting Abstracts}, vol.~2019 of \emph{APS
  Meeting Abstracts}, p.~Z10.002, Jan., 2019.

\bibitem{Digman:2019}
M.C.~Digman, C.V.~Cappiello, J.F.~Beacom, C.M.~Hirata and A.H.~Peter, \emph{Not
  as big as a barn: Upper bounds on dark matter-nucleus cross sections},
  \href{https://doi.org/10.1103/physrevd.100.063013}{\emph{Physical Review D}
  {\bfseries 100} (2019) }.

\bibitem{Farrar:2017eqq}
G.R.~Farrar, \emph{{Stable Sexaquark}},
  \href{https://arxiv.org/abs/1708.08951}{{\ttfamily 1708.08951}}.

\bibitem{Sakurai:QM}
J.J.~Sakurai and J.~Napolitano, \emph{{Modern quantum mechanics; 2nd ed.}},
  Addison-Wesley (2011).

\bibitem{Xu:2021cap}
X.~Xu and G.R.~Farrar, \emph{{Binding of Dark Matter with Nucleus: Capture
  Cross Section and Emission Signal}}, {\emph{(to appear)} (2021) }.

\bibitem{Khrapak:2004ieee}
S.A.~Khrapak, A.V.~Ivlev, G.E.~Morfill, S.K.~Zhdanov and H.M.~Thomas,
  \emph{Scattering in the attractive yukawa potential: application to the
  ion-drag force in complex plasmas},
  \href{https://doi.org/10.1109/TPS.2004.826073}{\emph{IEEE Transactions on
  Plasma Science} {\bfseries 32} (2004) 555}.

\bibitem{Loeb:2010gj}
A.~Loeb and N.~Weiner, \emph{{Cores in Dwarf Galaxies from Dark Matter with a
  Yukawa Potential}},
  \href{https://doi.org/10.1103/PhysRevLett.106.171302}{\emph{Phys. Rev. Lett.}
  {\bfseries 106} (2011) 171302}
  [\href{https://arxiv.org/abs/1011.6374}{{\ttfamily 1011.6374}}].

\bibitem{fStableS21inprep}
G.R.~Farrar, \emph{{A Stable Sexaquark: Overview and Discovery Strategies}},
  {\emph{in preparation} (2021) }.

\end{thebibliography}\endgroup
\bibliographystyle{JHEP}

\newpage

\appendix
\section{\label{sec:alphaAll}Exclusion on ($\alpha,m_X$) for all nuclei in the Dewar Experiment}

\begin{figure}
\centering 
\includegraphics[width=1.0\textwidth]{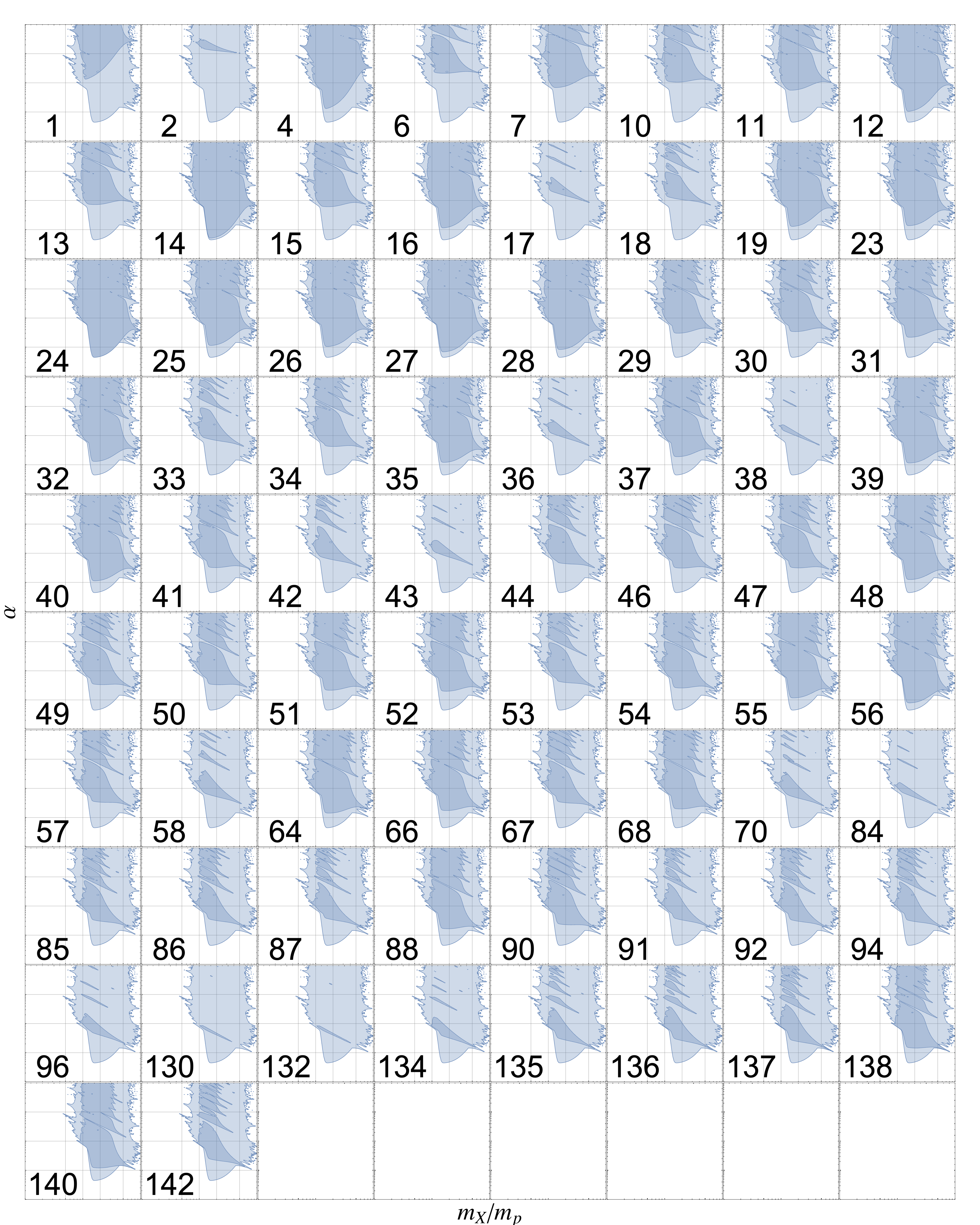}
\caption{\label{fig:alphaAllAttractive}Exclusion region on $(\alpha,m_X)$ and $m_\phi=1$ GeV for each $A$ for attractive interaction in the dewar experiment. The blue region are excluded from all the 74 different A combined and the darker region shows contribution from the labeled $A$. The axes are in Log scale with $m_X/m_p \in [0.1,100]$ and $\alpha \in [10^{-3},10]$. Grid lines indicate $m_X/m_p=(0.5,1,2,5)$ and $\alpha=(10^{-2},10^{-1},1)$.}
\end{figure}
\begin{figure}
\centering 
\includegraphics[width=1.0\textwidth]{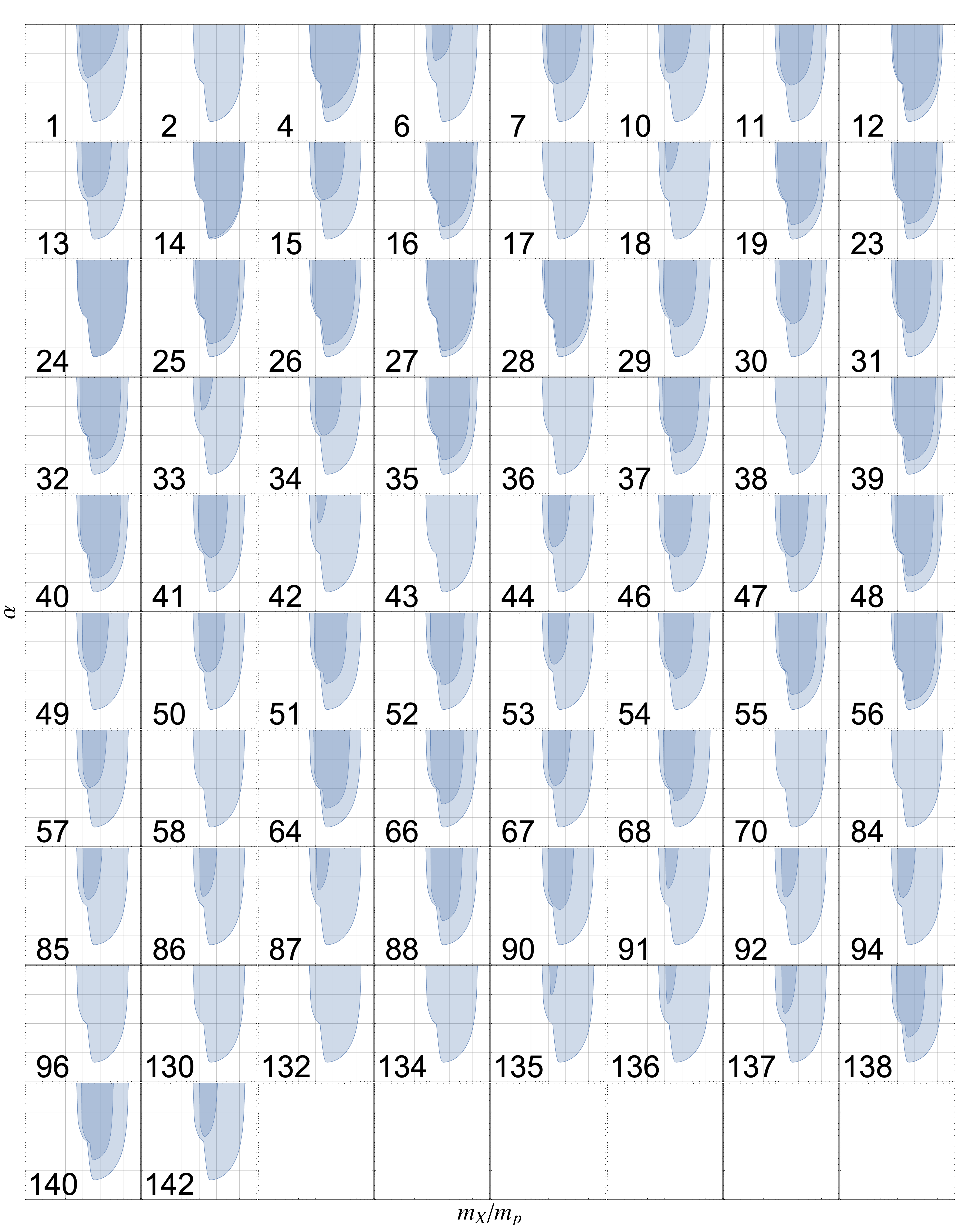}
\caption{\label{fig:alphaAllRepulsive}Exclusion region on $(\alpha,m_X)$ and $m_\phi=1$ GeV for each $A$ for repulsive interaction in the dewar experiment. The blue region are excluded from all the 74 different A combined and the darker region shows contribution from the labeled $A$. The axes are in Log scale with $m_X/m_p \in [0.1,100]$ and $\alpha \in [10^{-3},10]$. Grid lines indicate $m_X/m_p=(0.5,1,2,5)$ and $\alpha=(10^{-2},10^{-1},1)$.}
\end{figure}


\end{document}